\documentclass[aps,pre,twocolumn,showpacs]{revtex4}
\usepackage{amssymb,amsmath}
\usepackage[dvips]{graphicx}

\begin{document}

\title{Hole Structures in Nonlocally Coupled Noisy Phase Oscillators}

\author{Yoji Kawamura}
\email{kawamura@ton.scphys.kyoto-u.ac.jp}
%\email{ykawamura@jamstec.go.jp}
\affiliation{Department of Physics, Graduate School of Sciences,
Kyoto University, Kyoto 606-8502, Japan}
\affiliation{The Earth Simulator Center,
Japan Agency for Marine-Earth Science and Technology,
Yokohama 236-0001, Japan}

%\date{\today}
\date{December 15, 2006}

%\pacs{05.45.-a, 05.45.Xt, 82.40.Ck}
\pacs{05.45.Xt, 82.40.Ck}

%%%%% abstract
\begin{abstract}
  We demonstrate that a system of nonlocally coupled noisy phase
  oscillators can collectively exhibit a hole structure, which
  manifests itself in the spatial phase distribution of the
  oscillators.
  The phase model is described by a nonlinear Fokker-Planck equation,
  which can be reduced to the complex Ginzburg-Landau equation near
  the Hopf bifurcation point of the uniform solution.
  By numerical simulations, we show that the hole structure
  clearly appears in the space-dependent order parameter, which
  corresponds to the Nozaki-Bekki hole solution of the complex
  Ginzburg-Landau equation.
\end{abstract}

\maketitle

%%%%% section 1
%\section{Introduction} \label{sec:introduction}
One representative class of coupled oscillator systems is the coupled
{\it phase} oscillators
~\cite{ref:winfree80,ref:kuramoto84,ref:pikovsky01,ref:manrubia04}.
For example, {\it globally} coupled phase oscillators, such as the
{\it Kuramoto model}, have attracted the attention of many researchers
for a long time~\cite{ref:strogatz00,ref:kiss02,ref:acebron05}.
Recently, while coupled phase oscillators on {\it complex networks}
have been investigated widely
~\cite{ref:kori04,ref:ichinomiya04,ref:restrepo05,ref:boccaletti06},
{\it nonlocally} coupled phase oscillators have also been studied,
which exhibit a remarkable class of patterns called {\it chimera}
where phase-locked oscillators coexist with drifting ones
~\cite{ref:kuramoto02,ref:shima04,ref:abrams04,ref:kuramoto06,ref:kawamura07}.

%%The organization of the present paper is the following.
%%
In this paper, we demonstrate that nonlocally coupled
{\it noisy} phase oscillators can collectively exhibit
a hole structure in their phase distribution.
After briefly reviewing several results on the phase
model~\cite{ref:kuramoto06,ref:shiogai03}, we will present our new
findings obtained from numerical simulations of the Langevin-type
equation and its corresponding nonlinear Fokker-Planck equation
describing the phase model.
We will show that a hole structure clearly appears in a properly
defined order parameter under suitable conditions, and compare it with
the Nozaki-Bekki hole solution of the complex Ginzburg-Landau equation,
for which the modulus displays regions of local depression.

%%%%% section 2
%\section{Model} \label{sec:model}
A system of nonlocally coupled noisy phase oscillators is described
by the following Langevin-type equation (LE) for a phase $\phi(x,t)$
at location $x$ and time $t$:
%%% eq.1
\begin{equation}
\partial_t\phi=\omega+\int_{-\infty}^{\infty}dx'\,
G\left(x-x'\right)\Gamma\bigl(\phi(x,t)-\phi(x',t)\bigr)+\xi(x,t).
\label{eq:langevin}
\end{equation}
Here the first term $\omega$ represents the natural frequency common
to all the oscillators, the second term the nonlocal coupling, and the
last term the additive noise.
The phase coupling function $\Gamma(\phi)$, which is a $2\pi$-periodic
function of $\phi$, satisfies the in-phase condition, i.e.,
$d\Gamma(\phi)/d\phi|_{\phi=0}<0$~\cite{ref:kuramoto84}.
The spatial coupling function $G(x)$ is given by
%%% eq.2
\begin{equation}
G\left(x\right)=\frac{1}{2}\exp\left(-\left|x\right|\right),
\label{eq:nonlocal}
\end{equation}
which is normalized in the infinite domain.
The noise is assumed to be Gaussian-white, whose statistics are
specified by
%%% eq.3
\begin{equation}
\left\langle\xi\left(x,t\right)\right\rangle=0,\quad
\left\langle\xi\left(x,t\right)\xi\left(x',t'\right)\right\rangle
=2D\delta\left(x-x'\right)\delta\left(t-t'\right).
\label{eq:noise}
\end{equation}
Equation~(\ref{eq:langevin}) can be derived from a certain class of
reaction-diffusion systems under suitable conditions, using the phase
reduction method after adiabatically eliminating a highly diffusive
chemical component~\cite{ref:kuramoto06}.

In Refs.~\cite{ref:kuramoto06,ref:shiogai03}, it was shown that the
LE~(\ref{eq:langevin}) is equivalent to a single-oscillator nonlinear
Fokker-Planck equation (FPE) given by
%%% eq.4
\begin{equation}
\frac{\partial f(\psi,x,t)}{\partial t}=-\frac{\partial}{\partial\psi}
\Bigl[V\left(\psi,x,t\right)f\left(\psi,x,t\right)\Bigr]
+D\frac{\partial^2 f(\psi,x,t)}{\partial\psi^2},
\label{eq:fp}
\end{equation}
%%% eq.5
\begin{align}
V\left(\psi,x,t\right)=\omega
&+\int_{-\infty}^{\infty}dx'\,G\left(x-x'\right) \nonumber \\
&\times\int_0^{2\pi}d\psi'\,\Gamma\left(\psi-\psi'\right)
f\left(\psi',x',t\right),
\label{eq:drift}
\end{align}
where $f(\psi,x,t)$ is a space-time-dependent single phase
distribution function of $\psi$, i.e., the normalized probability
density that $\phi(x,t)$ takes a value $\psi$ (see also
Refs.~\cite{ref:risken89,ref:gardiner97}).

The phase model~(\ref{eq:langevin}) is capable of sustaining traveling
waves below a critical noise strength, $D = D_c$, where the uniform
solution of the FPE~(\ref{eq:fp}) undergoes a Hopf bifurcation
~\cite{ref:kuramoto06,ref:shiogai03}.
In the vicinity of this Hopf bifurcation point, we can derive a
complex Ginzburg-Landau equation (CGLE)
%%% eq.6
\begin{equation}
\partial_t A(x,t) = \lambda^2 \left(D_c-D\right) A
+ d \partial_x^2 A - g \left|A\right|^2 A,
\label{eq:cgl}
\end{equation}
from the FPE~(\ref{eq:fp}) by applying the center-manifold reduction
method~\cite{ref:kuramoto84}, where we introduced the complex
amplitude $A(x,t)$ representing the fluctuation of $f(\psi,x,t)$ in
the phase direction as
%%% eq.7
\begin{equation}
  f(\psi,x,t)=\frac{1}{2\pi}+\frac{1}{2\pi}
  \Bigl(A(x,t)e^{i\lambda\psi+i\Omega t}
  +A^{\ast}(x,t)e^{-i\lambda\psi-i\Omega t}\Bigr).
\label{eq:amp}
\end{equation}
Here $\lambda$ is the wavenumber of the phase fluctuation, $\Omega$
the Hopf frequency, and $A^\ast$ the complex conjugation of $A$.
The parameters of the CGLE~(\ref{eq:cgl}) are given by
%%% eq.8
\begin{equation}
D_c=\Im\Gamma_\lambda/\lambda,\quad
\Omega=-\lambda\left(\omega+\Re\Gamma_\lambda+\Gamma_0\right),
\label{eq:hopf}
\end{equation}
%%% eq.9
\begin{equation}
d=-i\lambda\Gamma_\lambda,\quad
g=\frac{\lambda\Gamma_\lambda
\left(\Gamma_{2\lambda}+\Gamma_{-\lambda}\right)}
{2\Im\Gamma_\lambda-i\Re\Gamma_\lambda+i\Gamma_{2\lambda}},
\label{eq:comp}
\end{equation}
where $\lambda=\arg\,\max_l\Im\Gamma_l/l$, and $\Gamma_l$ is the
Fourier component of the phase coupling function defined by
%%% eq.10
\begin{equation}
\Gamma(\psi)=\sum_{l=-\infty}^{\infty}\Gamma_l\,e^{il\psi}.
\label{eq:gammal}
\end{equation}
In what follows, we restrict ourselves to the case of $\lambda=1$ and
positive $\Re g$, i.e., the case that the first Fourier component of
the phase distribution function has the largest critical noise
strength, and the Hopf bifurcation of the uniform solution is
supercritical.

%%%%% section 3
%\section{Simulation} \label{sec:simulation}
Since we assume $\lambda=1$, Eq.~(\ref{eq:hopf}) and Eq.~(\ref{eq:comp})
depend only on the first and the second harmonics of the phase coupling
function, $\Gamma_{\pm 1}$ and $\Gamma_2$.
In this case, without loss of generality, the phase coupling function
can be expressed as
%%% eq.11
\begin{equation}
\Gamma\left(\psi\right)=
-\sin\left(\psi+\alpha\right)
+u\sin\left(2\psi+\beta\right).
\label{eq:gamma}
\end{equation}
Now let us introduce a space-time-dependent complex order parameter
with modulus $R(x,t)$ and phase $\Theta(x,t)$ as
%%% eq.12
\begin{align}
R\left(x,t\right)e^{i\Theta\left(x,t\right)}
&\equiv\int^{\infty}_{-\infty}dx'\,G\left(x-x'\right)e^{i\phi\left(x',t\right)} \nonumber \\
&=\int^{\infty}_{-\infty}dx'\,G\left(x-x'\right)\int^{2\pi}_{0}d\psi'\,
e^{i\psi'}f\left(\psi',x',t\right) \nonumber \\
&=\int^{\infty}_{-\infty}dx'\,G\left(x-x'\right)A^{\ast}\left(x',t\right)e^{-i\Omega t} \nonumber \\
&\simeq A^{\ast}\left(x,t\right)e^{-i\Omega t}.
\label{eq:order}
\end{align}
In deriving the last expression, we utilized the fact that the spatial
characteristic length of the complex amplitude becomes sufficiently
long compared to the nonlocal coupling length near the critical
point~\cite{ref:shiogai03,ref:kuramoto06}.
Thus, the order parameter corresponds to the complex conjugation of
the complex amplitude, which is governed by the CGLE~(\ref{eq:cgl}).
As is well known, CGLE admits the Nozaki-Bekki hole solutions
~\cite{ref:nozaki84,ref:lega84,ref:cross93,ref:mori97,ref:burguete99,ref:aranson02},
for which the existence of the amplitude degree of freedom is crucial.
In the following, we will show that the phase model~(\ref{eq:langevin})
can exhibit a hole structure in its space-time phase distribution,
despite its lack of apparent amplitude variables.

\begin{figure}
\centering
\includegraphics[height=5cm]{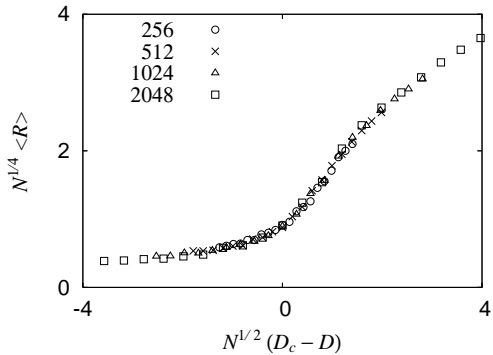}
\caption{Dependence of the order-parameter modulus on the noise strength,
  obtained from the Langevin-type equation~(\ref{eq:langevin})
  using different values of the number of oscillators $N$.
  The data are plotted using rescaled variables, revealing the scaling
  function given by Eq.~(\ref{eq:fss}).}
\label{fig:fss}
\end{figure}

%%% section 3.A
%\subsection{onset of coherence}
First of all, let us identify the Hopf bifurcation point predicted from
the FPE~(\ref{eq:fp}) by numerical simulations of the LE~(\ref{eq:langevin}).
In the numerical simulations, our continuous media of size $L$ is
replaced with a long array of $N$ oscillators with sufficiently small
separation $\varDelta x$ between the neighboring oscillators, i.e.,
$L=N\varDelta x$, and the periodic boundary condition is imposed.
In the continuous limit, $N \to \infty$ with $L$ fixed, the Langevin
simulation would exactly correspond to the Fokker-Planck simulation,
but we can use only finite values of $N$ in actual Langevin simulations.
Thus, we must consider the effect of finite-size fluctuations, which
comes from the finiteness of the oscillator number within the nonlocal
coupling range.
Applying the finite-size scaling argument developed in
Ref.~\cite{ref:pikovsky99} (see also Ref.~\cite{ref:kawamura04}) to
the space-time-averaged modulus $\langle R\rangle$ of the order
parameter, we can obtain a scaling form
%%% eq.13
\begin{equation}
N^{1/4}\left\langle R\right\rangle=F\left(N^{1/2}\left(D_c-D\right)\right),
\label{eq:fss}
\end{equation}
where $F$ is a scaling function depending on $N$ and $(D_c-D)$ only
through the combination $N^{1/2}(D_c-D)$.
We fix our system size $L$ to be 8.0 and vary the number $N$ of the
oscillators, i.e., $L=N\varDelta x=8.0$.
For this Langevin simulation, we fix the parameter values as
%%% eq.14
\begin{equation}
\alpha=0.5,\quad u=0.0,
\label{eq:para1}
\end{equation}
which yield the critical noise strength
%%% eq.15
\begin{equation}
D_c=\cos\left(0.5\right)/2,
\label{eq:dc-1}
\end{equation}
and the two essential parameters of the CGLE~\cite{ref:kuramoto84,ref:mori97},
%%% eq.16
\begin{equation}
c_1 \equiv \Im d / \Re d =  \tan\left(0.5\right),\quad
c_2 \equiv \Im g / \Re g = -\tan\left(0.5\right)/2.
\label{eq:c1c2-1}
\end{equation}
With these parameter values, spatially uniform oscillations are
realized in the Langevin simulation, as expected from the phase
diagram of the CGLE~\cite{ref:aranson02}.
Figure~\ref{fig:fss} summarizes the numerical results in the rescaled
variables, where $N^{1/4}\langle R\rangle$ is plotted as a function of
$N^{1/2}(D_c-D)$ for several values of $N$.
All curves collapse onto a single identical curve after rescaling,
which agrees well with the prediction based on the FPE~(\ref{eq:fp}).
This gives good evidence for the existence of the Hopf bifurcation
at $D = D_c$.
%%% fig.1

%%% section 3.B
%\subsection{hole structures}
Now we demonstrate that the phase model~(\ref{eq:langevin})
can exhibit a hole structure.
We first carry out a Fokker-Planck simulation.
The parameter values are chosen as
%%% eq.17
\begin{equation}
\alpha = \arctan\left(0.5\right),\quad
\beta = 2.3,\quad u=1.8,
%\beta=2.297606749396
\label{eq:para2}
\end{equation}
which give
%%% eq.18
\begin{equation}
D_c = \cos\left(\arctan\left(0.5\right)\right)/2,
\label{eq:dc-2}
\end{equation}
%%% eq.19
\begin{equation}
c_1 \equiv \Im d/\Re d = 0.5,\quad
c_2 \equiv \Im g/\Re g \simeq 2.0.
\label{eq:c1c2-2}
\end{equation}
The corresponding CGLE~(\ref{eq:cgl}) has stable hole solutions
with these parameter values~\cite{ref:aranson02}.
In our numerical simulations of the FPE~(\ref{eq:fp}), periodic
boundary conditions are imposed for both phase $\psi$ and space $x$.
The intervals of the phase and the space are $2\pi$ and $L=102.4$, respectively.
The pseudo-spectral method with $M=32$ modes is applied for the phase.
The number of the spatial grid points is $N=512$, i.e., $\varDelta x=0.2$.
Our numerical results are unchanged if we further increase the number
of modes $M$, the number of grid points $N$, or the system size $L$.
The initial condition is given by
%%% eq.20
\begin{equation}
f\left(\psi,x,t=0\right)=C_x\bigl[2+B(x)\exp(i\psi)+B^{\ast}(x)\exp(-i\psi)\bigr],
\label{eq:init1}
\end{equation}
where $C_x$ is a normalization constant, and
%%% eq.21
\begin{equation}
B(x)=
\begin{cases}
4(x - L / 4) / L & (0 \leq x < L / 2), \\
\exp\left[2\pi i(x - L / 2) / L\right] \; & (L / 2 \leq x < L).
\end{cases}
\label{eq:init2}
\end{equation}
Note that this $B(x)$ has one ``phase singularity'' and satisfies
the periodic boundary condition.
The noise intensity is chosen as $D/D_c=0.9$.

\begin{figure}
\centering
\includegraphics[height=4cm]{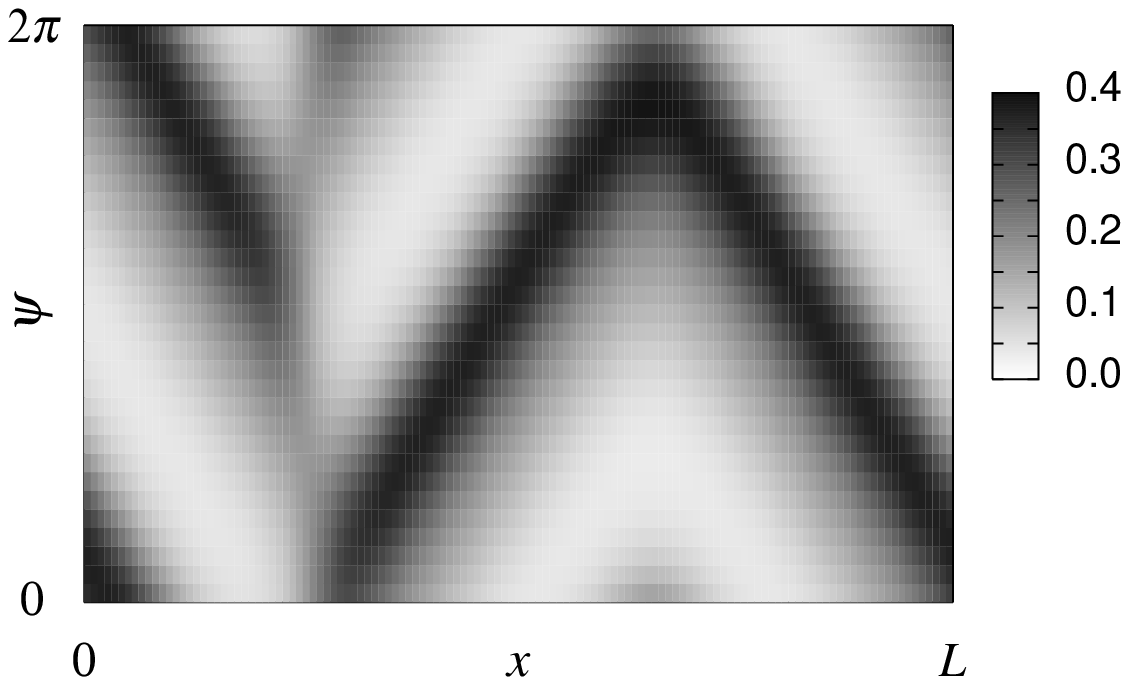}
\caption{Instantaneous spatial profile of the phase distribution
  function $f(\psi, x, t)$ obtained from a numerical simulation
  of the nonlinear Fokker-Planck equation~(\ref{eq:fp}).}
\label{fig:fp}
%\end{figure}
%%% fig.3
%\begin{figure}
%\centering
\includegraphics[height=5cm]{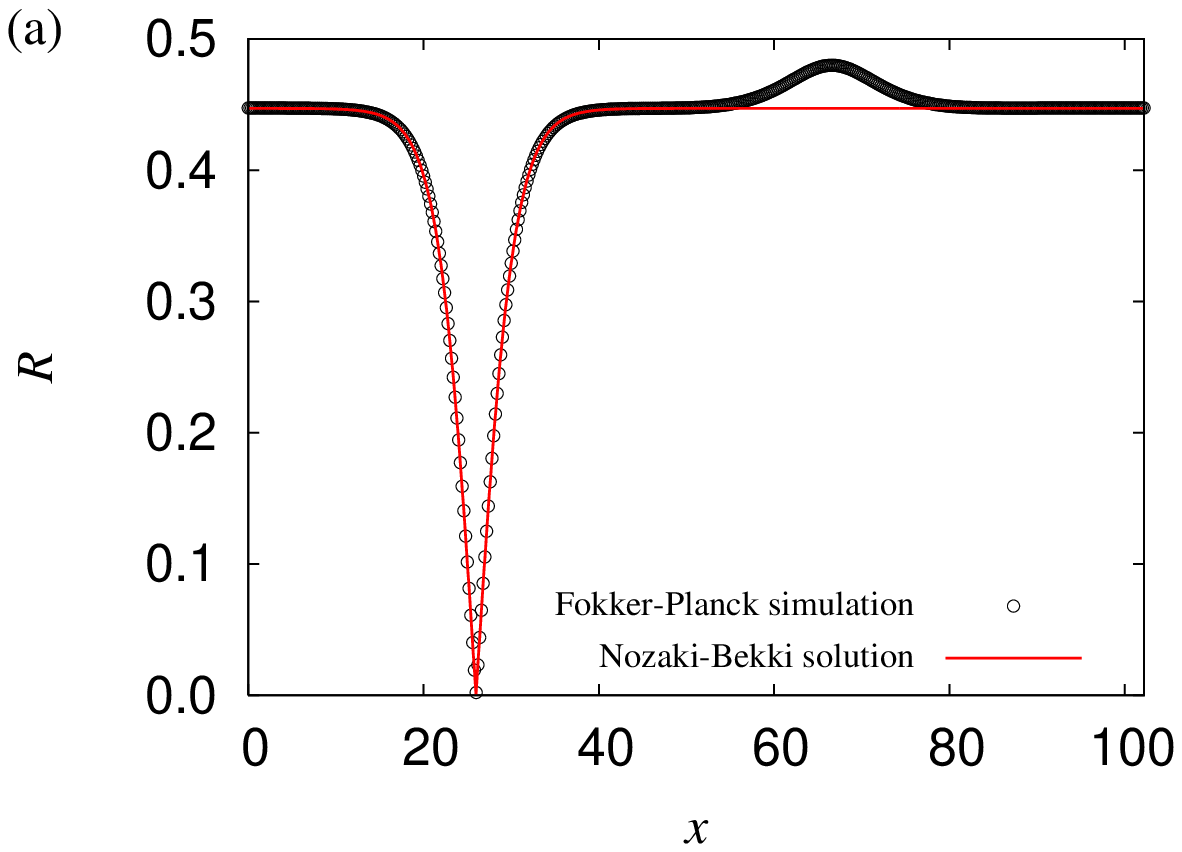}
\includegraphics[height=5cm]{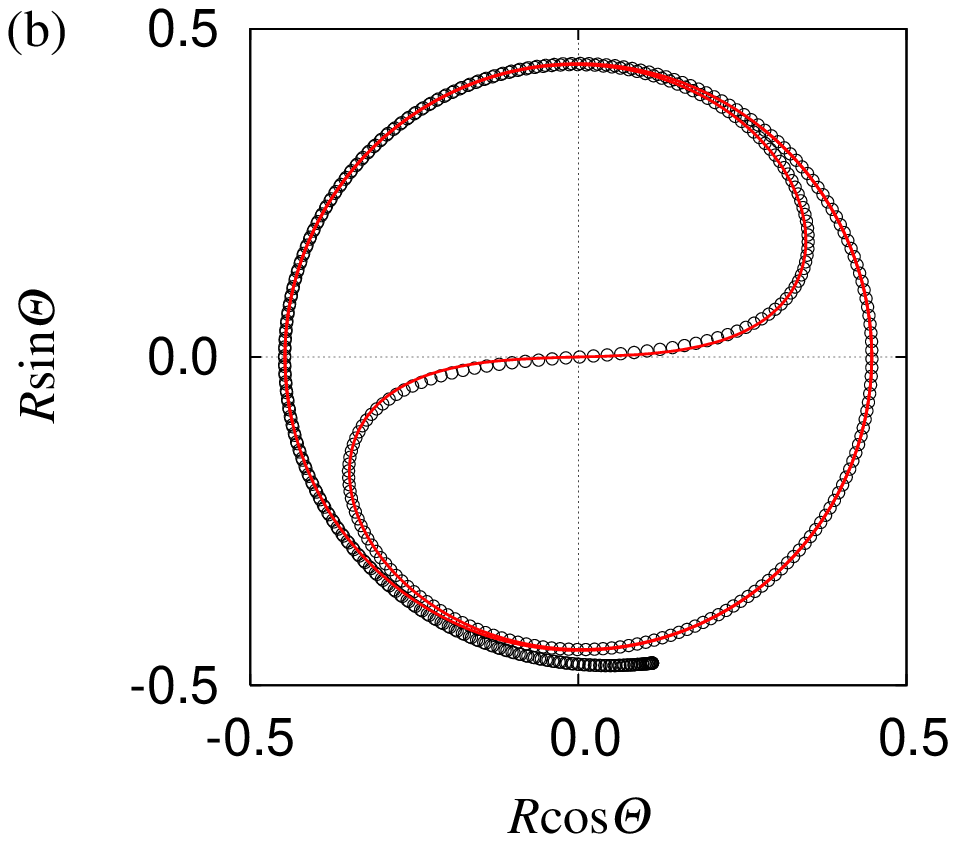}
\caption{(Color online)
  Instantaneous spatial profile of the order-parameter modulus (a).
  Instantaneous phase portrait of the order parameter (b).
  They are obtained from Fig.~\ref{fig:fp} using the relation 
  given in Eq.~(\ref{eq:order}).
  Solid lines are drawn using the Nozaki-Bekki hole solution of the CGLE
  given by Eq.~(\ref{eq:hole}) and Eq.~(\ref{eq:theta}), where the
  parameter values are estimated as $a_{\rm FP}\simeq 0.45$,
  $b_{\rm FP}\simeq 0.23$, $c_{\rm FP}\simeq 26$, and
  $s_{\rm FP}\simeq 0.15$.}
\label{fig:amp}
%\end{figure}
%%% fig.4
%\begin{figure}
%\centering
\includegraphics[height=4cm]{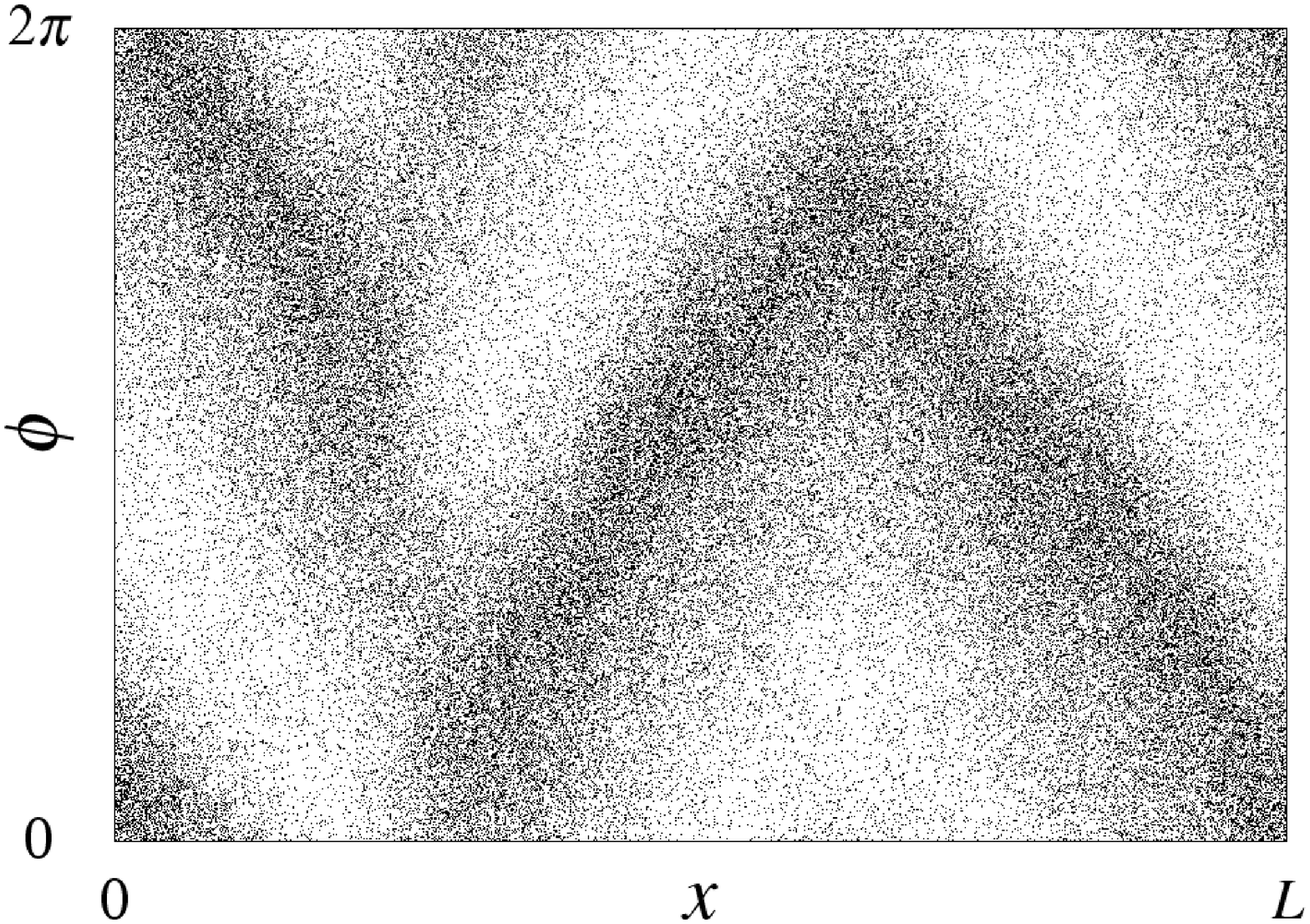}
\caption{Instantaneous spatial profile of the local oscillator
  phase obtained from a numerical simulation of the Langevin-type
  equation~(\ref{eq:langevin}).}
\label{fig:langevin}
\end{figure}

Figure~\ref{fig:fp} displays the spatial profile of the phase
distribution function $f(\psi,x,t)$ obtained from a numerical
simulation of the FPE~(\ref{eq:fp}).
Figure~\ref{fig:amp} displays the spatial profile of the
order-parameter modulus (a) and the phase portrait of the order
parameter (b), which are obtained from Fig.~\ref{fig:fp} using
the relation given in Eq.~(\ref{eq:order}).
We can confirm that a hole structure actually appears in the order
parameter.
The small bump at the right of the hole is a shock structure due to
the collision of counter-propagating plane waves emitted from the hole
structure.
Once such a hole structure is formed, it stably persists throughout
our numerical simulation.
This hole structure can be well fitted by a non-propagating Nozaki-Bekki
hole solution~\cite{ref:mori97} in the form
%%% eq.22
\begin{equation}
W_{\rm H}\left(x\right) =
a \tanh \left[b(x-c)\right] \exp \left[i\theta(x)\right],
\label{eq:hole}
\end{equation}
%%% eq.23
\begin{equation}
%\frac{d\theta}{dx} = s \tanh\left[b\left(x-c\right)\right],
d\theta / dx = s \tanh\left[b\left(x-c\right)\right],
\label{eq:theta}
\end{equation}
where $a$, $b$, $c$, and $s$ are real parameters.
These parameters are estimated as $a_{\rm FP}\simeq 0.45$,
$b_{\rm FP}\simeq 0.23$, and $s_{\rm FP}\simeq 0.15$ from the numerical
simulation of the FPE~(\ref{eq:fp}), while the theoretical values for
the reduced CGLE~(\ref{eq:cgl}) are given by $a_{\rm GL}\simeq 0.58$,
$b_{\rm GL}\simeq 0.27$, and $s_{\rm GL}\simeq 0.12$.
%%
%%The agreement between the simulation and the theory is about eighty
%%percents.
%%
The agreement between the simulation and the theory seems reasonable,
in consideration of various approximations used in deriving the
CGLE~(\ref{eq:cgl}) from the FPE~(\ref{eq:fp})~\cite{ref:comment1}.

We also carried out Langevin simulations with periodic boundary conditions.
We prepared an appropriate initial distribution of the oscillators
using the hole solution of the FPE~(\ref{eq:fp}).
Figure~\ref{fig:langevin} displays the snapshot of the
local oscillator phase obtained after the initial transient
for $L=102.4$ and $N=2^{17}$.
The spatial profile of the local oscillator phase well corresponds
to the spatial phase distribution shown in Fig.~\ref{fig:fp}.
However, this phase distribution corresponding to the hole structure
eventually collapses to that corresponding to a plane wave due to the
finite-size fluctuation that we mentioned above.
The lifetime of the hole structure clearly increases with $N$, so that
the hole structure is expected to exist stably in the $N \to \infty$
limit~\cite{ref:comment2}.
%%% fig.2

%%%%% section 4
%\section{Summary} \label{sec:summary}
In summary, we studied a system of nonlocally coupled noisy phase oscillators
based on the LE~(\ref{eq:langevin}) and its corresponding FPE~(\ref{eq:fp}).
We confirmed that the onset of the coherence in the order parameter of
the LE~(\ref{eq:langevin}) is identical to the Hopf bifurcation of the
FPE~(\ref{eq:fp}), using the finite-size scaling relation for the numerical
data obtained from the Langevin simulation.
We then demonstrated that a stable hole structure can appear in the
order parameter calculated from the FPE~(\ref{eq:fp}), which corresponds
to the non-propagating Nozaki-Bekki hole solution of the CGLE~(\ref{eq:cgl}).
Phase models generally lack the amplitude variables, which are crucial
for the hole solutions with phase singularities.
In the noisy phase model~(\ref{eq:langevin}), however, the external random
force effectively produces the amplitude degrees of freedom in the phase
distribution of the oscillators.
The phase model~(\ref{eq:langevin}) can exhibit yet another interesting phenomenon
called {\it noise-induced turbulence}~\cite{ref:kuramoto06,ref:shiogai03}.
Detailed numerical and theoretical analysis on this phenomenon
is recently reported in Ref~\cite{ref:kuramoto07}.

%%%%% acknowledgments
\begin{acknowledgments}
The author is grateful to Y.~Kuramoto, T.~Mizuguchi, H.~Nakao,
D.~Tanaka, and K.~Arai for useful discussions.
\end{acknowledgments}

%%%%% references


\begin{thebibliography}{99}
\bibitem{ref:comment1}
The main cause of the discrepancy is the CGLE approximation.
%%
The center-manifold reduction to the CGLE~(\ref{eq:cgl})
implicitly assumes that $\Re d$ and $\Re g$ are of the
same order~\cite{ref:kuramoto84,ref:mori97}.
%%
Our parameter condition~(\ref{eq:para2}) for a hole solution
gives $\Re d / \Re g \simeq 4$, which is rather large.
\bibitem{ref:comment2}
It is very difficult to precisely determine how the life time of
the hole structure increases with the number of oscillators $N$ by
numerical simulations of the Langevin-type equation~(\ref{eq:langevin}).
%%
To derive the dependence of the life time of the hole structure on $N$
is a challenging difficult task.
%%
For this purpose, investigations on a complex Ginzburg-Landau
equation with external noise would be more suitable.
\end{thebibliography}
\end{document}